\documentclass[preprint2]{aastex}
\pdfoutput=1

\shorttitle{Proper motion of the Arches cluster}
\shortauthors{A. Stolte et al.}

\begin{document}

\title{The proper motion of the Arches cluster with Keck Laser-Guide Star Adaptive Optics}
\author{A. Stolte\altaffilmark{1}, A. M. Ghez\altaffilmark{1,2}, M. Morris\altaffilmark{1}, J. R. Lu\altaffilmark{1}, W. Brandner\altaffilmark{1,3}, K. Matthews\altaffilmark{4}}
\altaffiltext{1}{Division of Astronomy and Astrophysics, UCLA, Los Angeles, CA 90095-1547, astolte@astro.ucla.edu, ghez@astro.ucla.edu, morris@astro.ucla.edu, jlu@astro.ucla.edu}
\altaffiltext{2}{Institute of Geophysics and Planetary Physics, UCLA, Los Angeles, CA 90095} 
\altaffiltext{3}{Max-Planck-Institut for Astronomy, K\"onigstuhl 17, 69117 Heidelberg, Germany, brandner@astro.ucla.edu}
\altaffiltext{4}{Caltech Optical Observatories, California Institute of Technology, MS 320-47, Pasadena, CA 91225, kym@caltech.edu}

\begin{abstract}
We present the first measurement of the proper motion of the 
young, compact Arches cluster near the Galactic center from 
near-infrared adaptive optics (AO) data 
taken with the recently commissioned 
laser-guide star (LGS) at the Keck 10-m telescope.
The excellent astrometric accuracy achieved with LGS-AO
provides the basis for a detailed comparison with
VLT/NAOS-CONICA data taken 4.3 years earlier. Over the 4.3 year 
baseline, a spatial displacement of the Arches cluster with respect 
to the field population is measured to be $24.0 \pm 2.2$ mas, 
corresponding to a proper motion of $5.6 \pm 0.5$ mas/yr 
or $212 \pm 29$ km/s at a distance of 8 kpc. In combination with 
the known line-of-sight velocity of the cluster, we derive a 3D space 
motion of $232 \pm 30$ km/s of the Arches relative to the field. 
The large proper motion of the Arches
cannot be explained with any of the closed orbital families
observed in gas clouds in the bar potential of the inner Galaxy, but 
would be consistent with the Arches being on a transitional trajectory between
x1 and x2 orbits. We investigate a cloud-cloud collision as the possible
origin for the Arches cluster.
The integration of the cluster orbit in the potential of the inner Galaxy 
suggests that the cluster passes within 10 pc of the supermassive black 
hole only if its true GC distance is very close to its projected distance.
A contribution of young stars from the Arches cluster to the young 
stellar population in the inner few parsecs of the GC thus appears 
increasingly unlikely. 
The measurement of the 3D velocity and orbital analysis provides the first 
observational evidence that Arches-like clusters do not spiral into the GC.
This confirms that no progenitor clusters to the nuclear cluster are 
observed at the present epoch.
\end{abstract}


\section{Introduction}

The Arches cluster is among the very few massive starburst
clusters observed in the inner Milky Way. At projected 
distances below 30 pc from the Galactic center (GC), only two 
other dense, young clusters are known: the Quintuplet 
and the central cluster. Obtaining an unbiased sample
of cluster members and thus an estimate of the stellar
mass function is difficult for the latter two clusters;
the central cluster contains a mixed variety of stars 
having different ages, and
the 4 Myr ``old'' Quintuplet cluster appears already
widely dispersed (see, e.g., Fig.~2 in Figer et al.~1999).
The compactness of the Arches cluster and its young, 
uniform age of only $\sim 2$ Myr (Najarro et al.~2004, 
Figer et al.~2002) characterize this cluster 
as a unique target to study the stellar mass function and
dynamical properties of clusters forming in the immediate
neighbourhood of the center of our Galaxy.  

The inner $\sim 100$ pc of the nuclear region of the 
Galaxy are a hostile environment for star cluster survival. 
Clusters like the Arches and Quintuplet are expected to disrupt 
in the intense tidal field on a time scale of $\sim 10$ Myr
(Kim et al.~1999, Portegies Zwart et al.~2002). The 
strong influence of external tidal forces on the 
evolution of these clusters led to the suggestion by Gerhard 
(2001) that massive, young clusters migrate inwards from 
tens of parsecs to the inner few parsecs around the black
hole. Gerhard suggested that the cores of these clusters might 
survive tidal disruption and supply 
the population of young, high-mass stars observed close 
to the black hole, where conditions are inhospitable to 
in-situ star formation (Morris 1993, Ghez et al.~2003, 2005). 
However, dynamical 
simulations suggest that a cluster mass of $M_{cl} > 10^6\,M_\odot$ 
is required for a dense core to reach the central parsec
within the lifetime of the high-mass stars (Kim \& Morris 2003),  
much more massive than the Arches today with $M_{cl}\sim 10^4\,M_\odot$.

The amount of material that was already stripped from the 
cluster and thus the initial cluster mass depend on its orbit
in the GC potential and the true distance of the cluster to 
the GC. At its projected distance of 26 pc, the Arches could
have lost 50\% of its initial mass, such that the 
cluster could have been as massive as $2\cdot 10^4\,M_\odot$ initially
(Kim et al.~2000).
Thus far, both the orbit and the true distance to the GC, $r_{GC}$,
are observationally unconstrained. Kim et al.~(2000) and Portegies
Zwart et al.~(2002) attempted to estimate $r_{GC}$ from N-body simulations 
of Arches-like clusters moving on circular orbits in a spherically 
symmetric potential representing the GC tidal field.
Both studies compared the surface density profiles of dynamically
evolving clusters to the surface density of massive stars observed 
in the Arches. Despite similar initial conditions (cluster mass, age, 
Galactic potential), the most likely GC distance ranges 
from 30-50 pc, favoring 30 pc \citep{kim2000}, to 50-90 pc, 
favoring 90 pc \citep{pz2002}.
The measurement of the 3-dimensional space motion of the Arches would 
provide a strong observational constraint on the cluster orbit.
An estimate of the true distance from the GC and of the orbital 
timescale limits the amount of material stripped during the 
cluster's lifetime, and thus constrains the initial mass
of the Arches and the galactocentric radius where the
complete tidal dissolution of the cluster is expected.

In this paper, we present an astrometric comparison of
new laser guide star adaptive optics (LGS-AO) observations
obtained at Keck with natural guide star (NGS) AO data 
observed 4.3 years earlier with the Very Large Telescope (VLT).
The high spatial resolution and astrometric accuracy of the
AO observations allow us to derive the proper motion of the 
Arches cluster with respect to the field population.
In combination with the known line-of-sight velocity, the 3D space motion 
of the cluster provides the first observational constraints on 
the cluster orbit.  

We present our observations, data reduction and photometric 
calibration in Sec.~\ref{obssec}.
The critical step of astrometric matching between the NIRC2 and NACO 
data sets is explained in Sec.~\ref{geosec}. The proper motion is 
derived in Sec.~\ref{pmsec}, and a discussion of the 3D space motion 
of the cluster is presented in Sec.~\ref{orbsec}. The 3D orbit of the
cluster in the GC potential is modelled in Sec.~\ref{orbitsec},
followed by our conclusions in Sec.~\ref{sumsec}.

\section{Observations}
\label{obssec}

\subsection{Keck/NIRC2}
\label{kecksec}

KeckII LGS-AO observations were obtained on 2006 July 18.
The NIRC2 narrow camera with a pixel scale of 
10 mas/pixel ($0.01^{''}$/pix) was employed to 
obtain a set of 50 images of the Arches cluster
in $K'$ ($\lambda_c = 2.12 \mu$m, $\Delta\lambda = 0.35\mu$m).
The laser guide probe provided a guide source with 
$V\sim10$ mag brightness centered in the field of view
of the NIRC2 camera. 
A foreground giant with $R = 15.2$ mag located $10^{''}$ from 
the field center was used as the tip-tilt reference source. 
The 5 mag brighter LGS -- as opposed to the faint tip-tilt reference 
source previously available as a natural guide star (NGS) -- delivered 
images very close to the diffraction limit of 52 mas at $2.12\mu$m.
These observations exceed the resolution obtained with earlier NGS observations
\citep{stolte2005}
by a factor of 1.5, which facilitates resolving the stellar population
in the crowding limited cluster field.  
To aid reconstruction of bad pixels in NIRC2 and avoid 
detector persistence from bright sources, the telescope
was dithered in a random dither pattern with 3 images taken at 
each position, with individual integration 
times of 30s per frame, split into 10 coadds $\times$ 3s
in the $K'$ filter to avoid saturation of most bright sources.
The visual seeing during these observations was 
$0.\!^{''}8 - 1.\!^{''}2$, which is worse than average for
Mauna Kea conditions, leading to larger variations than
expected in AO performance. 
Thirty-six $K'$-band frames displaying the best AO 
performance were selected, leading to a total exposure 
time of 18 minutes in $K'$.  These data form the basis 
of the proper motion study.

In addition, 15 images in $H$ 
($\lambda_c = 1.63 \mu$m, $\Delta\lambda = 0.30 \mu$m) 
were obtained on 2006 May 21 with the same observational
setup. The fainter source brightness in $H$ allowed for longer
integration times of 10s $\times$ 3 coadds, and a box five 
dither pattern was used with 3 images taken at each position. 
Varying AO performance 
led to the selection of the 9 images displaying 
core FWHM $< 65$ mas, with a total exposure time 
of 4.5 minutes. 
The $H$-band data are used supplementary to the $K'$ 
images to reject artefacts from the final catalogue.

The standard imaging reduction steps including 
flat fielding, sky and dark subtraction, and 
removal of cosmic ray events are detailed below. 
The science frames were reduced using flat fields
and offset sky frames taken during the same night.
Individual bad pixel masks were created 
from a combination of dark, flat field, and cosmic ray 
masking of individual exposures. Strehl ratios on 
pre-selected $K'$ frames ranged from 0.25 to 
0.44 (median 0.34) corresponding to a PSF core FWHM
between 52 and 63 mas (median 56 mas).
$H$-band Strehls covered 0.15 to 0.19 (median 0.18) with 
a PSF core FWHM of 58 to 65 mas (median 60 mas).
In order to enhance the resolution in the 
final image, each frame was weighted
by the Strehl ratio prior to image combination. 
The individual bad pixel mask was applied to each 
frame, and frames were averaged together using
the drizzle algorithm (Fruchter \& Hook 2002).
The combined images display a core 
FWHM of 53 mas in $K'$ and 60 mas in $H$.
In addition to the deep $K'$ frame, three auxiliary $K'$ frames
were formed by coadding 12 individual images in each frame.
These auxiliary frames were used for a first pass rejection 
of image and PSF artefacts, as well as for uncertainty estimates.

Stellar positions and photometry were extracted from 
the final, averaged NIRC2 images using the Starfinder code
(Diolaiti et al.~2000). This code has the advantage
of constructing the PSF from the image empirically, 
without the need to assume an analytic function.
This is particularly critical for the complex AO
PSFs displaying steep kernels with structured 
haloes and Airy ring patterns. In the crowded
stellar field of the Arches, 10 stars were used to 
create the PSF. The detection threshold was set to 
3$\sigma$ above the local background, and a PSF correlation
threshold of 0.7 was imposed. The relatively low correlation threshold
enables us to include close neighbours in the final photometry 
file, while higher thresholds tend to omit faint sources 
embedded in the haloes of bright stars.
In order to reject false detections, each star entering the calibrated
$K'$ catalogue was required to be detected in the deep $K'$ exposure and in
all three auxiliary frames. This strict requirement led to the 
rejection of most PSF and image artefacts. Any remaining PSF artefacts 
were removed during the final catalogue matching. The calibrated 
NIRC2 $K'$ catalogue contained 869 objects in the magnitude range 
$10.3 < K' < 20.5$ mag on the combined
$10.\!^{''}8 \times 10.\!^{''}8$ image.
Photometric calibration was performed by zeropointing against 
the NACO photometry presented in Stolte et al.~(2005), 
and the comparison of $(H-K')_{NIRC2}$ 
with $(H-K_s)_{NACO}$ yielded no color terms.
A detailed photometric analysis of the cluster population 
will follow in a subsequent paper.

\subsection{VLT/NAOS-CONICA}
\label{nacosec}

\begin{figure*}[ht]
\includegraphics[width=8cm]{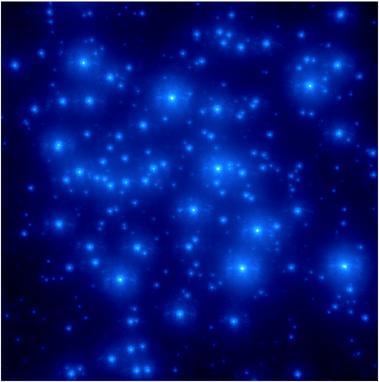}
\includegraphics[width=8cm]{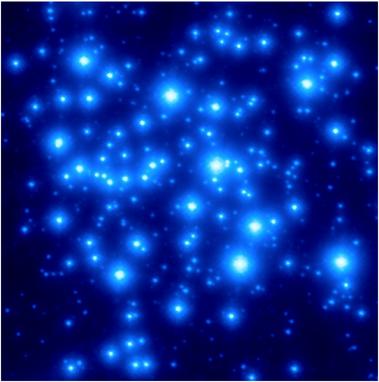}
\caption{\label{pics}
The LGS-AO Keck/NIRC2 $K'$ image (left) with a spatial resolution 
of 53 mas (FWHM PSF) compared to the NGS-AO VLT/NACO $K_s$ image
with 84 mas (FWHM PSF). The Keck 10 m diffraction limited image
displays the first order airy rings around each star.
North is up and East to the left.}
\end{figure*}

The Keck/NIRC2 $K'$ astrometric data set was combined with 
VLT/NACO data taken on 2002 March 31, 
providing a baseline of 4.3 years for the proper motion 
study. NACO $H$ and $K_s$ images with a total integration time
of 14 and 7 min, respectively, were obtained with the medium camera 
with a platescale of 27 mas/pix. The NACO data were taken with a  
natural guide star (NGS) on the NAOS optical wavefront sensor.
Because of the faintness of the NGS in the 
Arches field ($R = 15.2$ mag), the Strehl ratios 
achieved were 0.20 in $K_s$ and 0.14 in $H$, providing a FWHM 
resolution of 84 mas in both filters. The merged catalogue 
of sources detected in both $H$ and $K_s$ contains 487 stars in 
the magnitude range $10.2 < K_s < 20.7$ mag, and provides the 
basis for the proper motion measurement. 
A comparison between the NIRC2 $K'$ and NACO $K_s$ images is 
shown in Fig.~\ref{pics}. 
For a description of the NACO data and analysis
see Stolte et al.~(2005). 

\subsection{Geometric transformation and astrometry}
\label{geosec}

With the aim of measuring the relative motion of stars
in the Arches field, the spatial displacement of each
star over the 4.3 year baseline is calculated. 
Bright sources on the combined NIRC2 $K'$ image provide 
the reference frame for the proper motion analysis. 
A geometric transformation to convert NACO pixel positions 
to the NIRC2 frame was derived using the IRAF 
{\sl geomap} task. By deriving the transformation 
from bright and intermediate sources measured 
directly on both images, the different optical 
distortions between NIRC2 and NACO are taken 
into account within the limitations of the 
two-dimensional, second-order polynomial fit performed with 
{\sl geomap}. The geometric solution is created 
iteratively by minimizing the x,y residual displacements.
The final residual rms deviation was 
2 and 0.7 mas in the x and y direction, respectively.
In the following analysis, we therefore assume a systematic 
contribution of 2 mas in the proper motion uncertainty. 

The selection of 
proper motion members was based on matching the 
NIRC2 source catalogue with the transformed NACO 
catalogue using a matching radius of 40 mas. 
This radius provided the detection of 
most moving sources while avoiding wrong matches
with close neighbours in the dense stellar cluster 
field. In order to exclude artefacts from the 
complex PSF structure, we require that each source
be detected in either the NACO or NIRC2 $H$-band 
image in addition to both $K$-band images.
After blind matching, the matched sources were
visually inspected on the NIRC2 and NACO $K$-band
images, and 13 sources with merged centroids in NACO
due to the lower resolution or located on the edge of the 
$10^{''}$ NIRC2 field of view were removed from 
the matched catalogue, as the biased definition of 
their stellar centroids affected their apparent motion.
The final proper motion catalogue comprises 419 objects.

\section{Results} 
\label{pmsec}

\subsection{The relative motion of the Arches to the field}
\label{relpmsec}

The proper motion of the Arches cluster is derived from 
the relative motion of the cluster with respect to field
stars. This relative motion depends on the geometric 
transformation between both data sets. As detailed below,
the transformation is chosen such that proper motions over 
the 4.3 yr baseline are measured in the reference frame 
of the cluster. Before deriving the proper motion of the 
cluster, this reference frame will be introduced.

\begin{figure}[t]
\hspace{-1cm}
\includegraphics[width=6.4cm, angle=90]{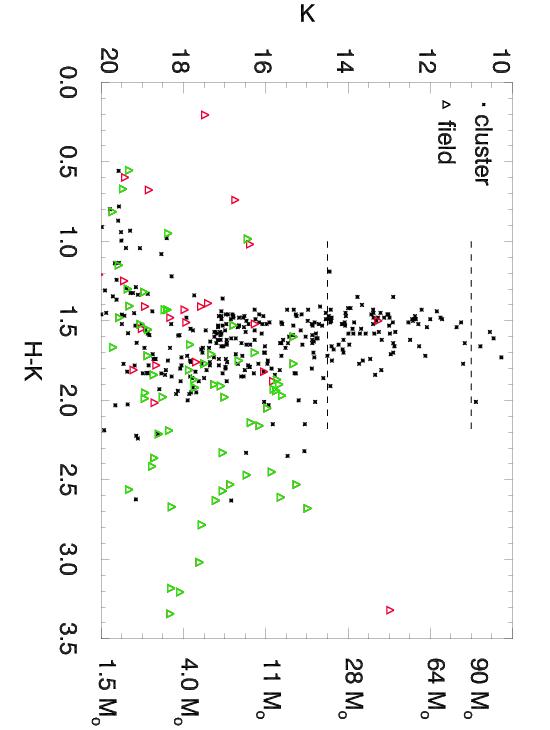}
\caption{\label{cmd}
Color-magnitude diagram of the central Arches cluster ($r < 0.25$ pc). 
Proper-motion selected cluster candidates (black asterisks, Fig.~\ref{xymove}) 
predominantly fill the Arches main sequence, while likely field stars 
(triangles) display redder colors and a larger scatter. Green triangles 
denote reference field stars used to derive the cluster proper motion, 
while red triangles are foreground candidates with irregular motions.
The bright stars on the Arches main sequence (dashed lines) define 
the geometric transformation between the NACO and NIRC2 $K$-band images.
For the faintest stars, the deeper NACO $H$-band magnitudes are matched
with NIRC2 $K'$. The mismatch in resolution between the two data sets 
introduces additional scatter.
The mass conversion for a 2 Myr Geneva main sequence isochrone 
\citep{lejeune2001} representing the cluster is 
indicated for reference on the right axis.}
\end{figure}

\begin{figure}[t]
\hspace*{-1.5cm}
\includegraphics[width=8.4cm,angle=90]{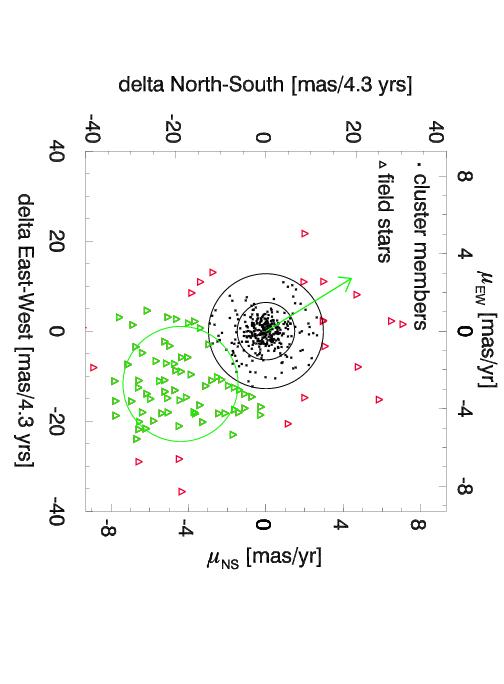}
\caption{\label{xymove}
Spatial displacement in the East-West and North-South directions
over 4.3 years in the cluster reference frame. 
The top and right axes display the corresponding proper motion.
Arches candidate members cluster around (0,0) (dots), while likely 
field stars (triangles) scatter to relative motions as large as 
40 mas/4.3 yrs or 9.3 mas/yr. The $1\sigma$ and $2\sigma$
error circles from a Gaussian fit to cluster candidate
motions are shown as black circles, and stars with proper motions 
less than $2\sigma$ are selected as member candidates. 
The bulk of the field stars are 
located in the South-West quadrant, indicating 
that the cluster moves to the North-East with respect
to the field population. 
The mean relative velocity of field stars (green triangles) yields 
a cluster motion of $212 \pm 29$ km/s to the North-East.
The green circle denotes the expected proper motion regime
of field stars moving on circular orbits in the bulge potential,
as derived from the line-of-sight velocity boundaries in Fig.~\ref{radvel}.
Sources marked in red were excluded due to their large deviation from the 
mean of the field population.}
\end{figure}

The geometric transformation between NACO and NIRC2 was derived from
bright sources in the magnitude range $11 < K_s < 14.5$ mag.
As can be seen in the color-magnitude diagram (CMD, Fig.~\ref{cmd}), 
this regime is composed of stars 
on the Arches main sequence and thus likely cluster members.
By defining the transformation from cluster stars, a
cluster reference frame is created in which
Arches members cluster around an East-West motion of $\Delta {\rm EW}=0$, 
and North-South motion $\Delta {\rm NS}=0$, implying that in the cluster 
reference frame the net spatial displacement of the 
cluster should be zero.
The spatial displacement of stars in the Arches field is 
shown in East-West and North-South motion
in Fig.~\ref{xymove}.
Two distinct features dominate in this diagram:
i) the dense clump of stars clustering around (0,0), 
which are cluster member candidates in the cluster 
reference frame, and ii) the scattered population of 
sources in the lower right quadrant. 

The fact that Fig.~\ref{xymove} reveals two distinct 
populations provides a means of identifying Arches 
cluster members based on proper motions. 
The $1\sigma$ and $2\sigma$ circles from a gaussian fit
to the relative motion distribution centered on zero
are enveloping member candidates in Fig.~\ref{xymove},
and Arches members are selected with less than $2\sigma = 12.8$ mas
spatial displacement in 4.3 years, corresponding to a proper 
motion $\le 3.0$ mas/yr. 
While cluster candidates are confined to 
a dense clump with small intrinsic scatter ($\sigma_{clus} = 0.73$ mas/yr (rms)),
the field population displays twice as large rms scatter
($\sigma_{field} = 1.6$ mas/yr (rms)). This scatter is expected if we observe 
a mixed population of stars tangentially or radially
along their orbits over a large range of distances.
Field stars along the line of sight are expected to display a 
wide range in dynamical properties depending on their orbits. 
It is therefore surprising that field stars are predominantly 
found in the South-West quadrant in Fig.~\ref{xymove}. 
This apparent South-West displacement of field stars in the 
Arches reference frame can most easily be explained by
the Arches moving to the North-East with respect to the 
field. The observed North-East motion of the cluster 
is consistent with a recent prediction 
from the X-ray emission of the bow shock created by the 
Arches during its impact on the nearby cloud \citep{wang2006}.
The median displacement of the field in the cluster 
reference frame is $24.0 \pm 2.2$ mas/4.3 years or 
$5.6 \pm 0.5$ mas/yr, which translates to a relative
motion of the cluster with respect to the field of
$\mu_{Arches} = 212 \pm 20$ km/s assuming a distance of 8 kpc. 
With a position angle (PA) of $31.\!^{\rm o}7 \pm 3.\!^{\rm o}8$ as measured from 
North to East, the direction of the proper motion vector is  
parallel to the Galactic plane (PA$=34.\!^{\rm o}8$) within the uncertainties.

Before employing the proper motion to derive
the 3-dimensional space motion of the cluster, we investigate
the sources of uncertainty contributing to the proper motion measurement.
The quoted accuracy in the proper motion measurement is
achieved by the large number of 67 reference field sources
contributing to the average relative motion between the cluster
and the field. 
The uncertainties were calculated adding the fixed component from 
the geometric transformation, 2 mas/4.3 yrs = 0.5 mas/yr, in
quadrature to the standard deviation in field star velocities,
$\sigma_{field}=1.6$ mas/yr, normalized to the number of field stars, 
$N = 67$, $\sigma_{field}/\sqrt{N} = 0.2$ mas/yr. The uncertainty 
from the zeropoint defined by the 328 cluster candidates is
$0.73/\sqrt{328} = 0.040$ mas/yr, and therefore negligible.
The final uncertainty from random deviations in the relative motion of 
$\sigma_\mu = \sqrt{0.5^2 + 0.2^2} = 0.54$ mas/yr is 
dominated by the geometric transformation between NACO 
and NIRC2, which includes residual optical distortion 
differences and the resolution-dependent 
uncertainty from fitting the centroid to the light 
distribution of each star. 

In addition, one might consider 
that the internal velocity distribution may contribute to 
the rms scatter in cluster candidate motions.
The cluster velocity dispersion can be estimated as 
$\sigma = \sqrt{0.4GM_{cl}/r_{hm}}$ (Binney \& Tremaine 1987, 
eq. 4-80b), where $M_{cl} = 10^4\,M_\odot$ is the total cluster mass 
and $r_{hm} = 0.4$ pc the half-mass radius \citep{stolte2002}. 
This yields $\sigma = 7$ km/s or 0.2 mas/yr, which is significantly
smaller than the rms scatter of 0.73 mas/yr 
observed in candidate cluster stars. 
The systematic uncertainty of 2 mas/4.3 yrs 
or 0.5 mas/yr imposed from the geometric transformation
between NACO and NIRC2 presently limits our ability to detect
the internal velocity dispersion of the cluster.
With a relative astrometric accuracy of 0.3 mas measured between 
the three NIRC2 auxiliary frames, the internal velocity dispersion
can be observed on a timescale of a few years.
A second epoch of NIRC2 data will avoid the transformation 
uncertainty inherited from the different optical distortions 
and spatial resolution in the NACO and NIRC2 data sets, thereby 
decreasing the overall uncertainty in the proper motion measurement,
and providing the potential to resolve the internal velocity dispersion
of the cluster.

\subsection{Sources of systematic uncertainty}

The random uncertainty derived above dominates the proper motion 
measurement only in the absence of systematic uncertainties.
These uncertainties can arise from the choice of selected field 
stars, as well as a bias towards field sources on the near side 
of the bulge.
In Fig.~\ref{xymove}, the field selection was based on the 
density of stars in proper motion space. We can investigate the 
uncertainty in our measurement due to the field sample selection
using defined subsamples. As we expect a larger positional 
uncertainty for fainter stars, the first sample consists of 
stars brighter than the apparent cluster main sequence cutoff
at $K=17.4$ mag. As bulge stars are predominantly faint, this 
bright sample consists of 27 sources, and yields $\mu = 221 \pm 22$ km/s.

As a consequence of the high foreground extinction towards the 
Galactic center, our observations could be biased to stars on 
the near side of the bulge. Such a bias should be revealed
in a proper motion difference between blue and red sources.
Thus we define two sets for our second sample: a set blueward
of $H-K = 2.1$ mag, encompassing the cluster main sequence and 
the bulge red clump, and a very red sample $H-K > 2.1$ mag with 
presumed background stars. The cluster proper motion derived from the 
``blue'' field sample of 45 stars becomes $194 \pm 19$ km/s, while 
the red sample of only 22 stars yields $249 \pm 22$ km/s.
The larger number of stars in the ``blue'' sample indicates 
that there is a bias towards stars on the near side of the bulge
in the complete sample. 
The difference of $55 \pm 29$ km/s is in 
excellent agreement with the proper motion difference of stars on 
the near and far side of the bulge, measured using red clump giants 
to $1.5 \pm 0.1$ mas/yr or $57 \pm 4$ km/s \citep{sumi2003}.
The higher cluster motion of $249 \pm 22$ km/s as measured from 
the very red sources indicates that the bias acts to decrease 
the cluster velocity. The absolute proper motion might be closer
to the mean between the red and blue sample, $\sim 221 \pm 29$ km/s,
in good agreement with the $212 \pm 20$ km/s derived for the full sample.
Nevertheless, the lower number of stars in the red sample indicates 
that sources with high extinctions or large distances, which have 
higher relative streaming velocities than sources closer to the GC, 
are predominantly lost. As this problem arises in all proper motion 
studies toward the GC, the maximum expected velocity for stars on the 
far side of the bar is not firmly known. This implies that the proper 
motion measurement remains a lower limit.

A final concern may arise from the sampling of field stars itself.
The bias from the $n=67$ field star sample can be estimated 
by deriving the cluster motion from a random sample of half of 
the field stars, $n=34$. In 100 random samples we obtain 
proper motion values between 195 and 227 km/s, with a mean and 
standard deviation of $211 \pm 20$ km/s, in excellent agreement 
with the value derived from the full sample above. In summary, 
we conclude that the only systematic uncertainty in the proper 
motion might arise from the bias to bulge stars in front of the GC,
but the effect is small and the measurements agree very well 
within the uncertainties.
Therefore, we use the full sample of field stars to obtain a 
conservative lower limit to the proper motion of 
the cluster in an absolute reference frame, and adjust the uncertainty
to account for the possible bias accordingly, $212 \pm 29$ km/s.

\subsection{The 3D motion of the Arches in the bulge reference frame}

In order to derive the absolute motion of the Arches cluster 
in the Galaxy, we need to understand the field restframe.
The CMD (Fig.~\ref{cmd}) shows that most of the proper motion 
selected field stars are faint and red. The clump
of field stars at $K=16, H-K=1.9$ mag is consistent with red clump 
stars at $d=8$ kpc and $A_K=3$ mag
(${\rm M_K}=-1.6$, Alves 2000), and the faint population, $K>17$ mag,
covers the regime of evolved bulge stars. Sumi et al.~(2003) analysed
a sample of $\sim\,50,000$ stars repeatedly observed in OGLE-II fields 
towards the bulge, and concluded that proper motions of red giants
and red clump stars are consistent with zero mean proper motion 
with respect to the GC. Only blue foreground disk
stars display a systematic deviation in mean proper motion from 
the GC, which is interpreted as the tangential motion 
of disk stars at 1-2 kpc distance from the Sun. In the NIRC2 field 
of view, we only observe seven stars significantly bluer ($H-K < 1.3$ mag)
than the deeply extincted cluster population 
($A_V\sim 24$ mag, Stolte et al.~2002), 
so that we can safely rule out significant contamination by disk stars. 

Sumi et al.~(2003) derived the streaming motion of red clump
stars in bulge fields from proper motions, and interprete the 
observed velocity difference of 57 km/s between sources in front 
of and behind the GC as
the rotation of the galactic bar in the same direction around the
GC as the Sun. This streaming motion is consistent with 
the bar models constructed from CO line-of-sight velocity maps 
(Binney et al.~1991, Englmaier \& Gerhard 1999).
In the context of the 3D reference frame, a detection bias 
towards sources on the near side of the bar causes a 
systematic loss of stars with proper motions in the 
opposite direction to the Arches motion
(compare to the orbit sketches in Fig.~\ref{orbits}). 
The number counts of 22 very red sources as opposed to 
45 blue sources indicate a loss of 
sources at the far side of the bar. The selective 
detection of sources in the front of the GC shifts the 
apparent mean motion of the field stars away from zero
in the same direction as the Arches cluster, which, when corrected, 
increases the cluster velocity with respect to the GC 
even further. The 3D motion derived from the proper 
motion sample is therefore a lower limit to the 
space motion of the Arches with respect to the GC.
\footnote{Although we cannot fully exclude
the possibility that the cluster is observed against a dense 
comoving group in the background, it appears very unlikely that 
the majority of field sources would be part of such a group.}

The heliocentric line-of-sight velocity of the Arches cluster 
was measured spectroscopically to be $+95\,\pm\,8$ km/s \cite{figer2002}.
Combining the proper motion with the line-of-sight velocity
yields the relative 3-dimensional space motion 
of the Arches, $v_{3D} = 232 \pm 30$ km/s, directed away
from the GC to the North-East, and away from the Sun. 

\section{Discussion and orbit considerations}
\label{orbsec}

The measurement of the Arches space velocity allows us 
to compare the cluster velocity with measured orbital velocities
of stars and clouds in the inner Galaxy, with the aim of
deriving the first qualitative constraints on the cluster orbit around the GC
by considering known orbital families in comparison with the 3D cluster motion.
The cluster orbit and, in particular, the GC
distance are critical for the formation and dynamical evolution 
of the cluster in the GC potential, for the amount of tidal mass 
loss of cluster members along its orbit, and for the potential 
for inspiral towards the GC. 

\subsection{Excluding a circular orbit?}
\label{circsec}

Due to the earlier lack of knowledge on the cluster motion
and the limited understanding of the inner Galaxy, and in particular
the ignoring of the thick nuclear stellar disk between 20 and 200 pc (Launhardt et al.~2002),
previous studies of the dynamical evolution and tidal disruption 
of the Arches in the GC potential assumed for simplicity that the cluster 
moves on a circular orbit in a spherically symmetric mass distribution
(Kim et al.~2000, Portegies Zwart et al.~2002). 
Without further constraints, a circular orbit in 
a spherically symmetric potential is the simplest case and most 
straightforward assumption. From this assumption, the amount of 
mass loss and the true distance of the Arches from the GC were 
derived. 

In the case of a circular orbit in a spherically symmetric potential,
the orbital parameters are determined by
the 3D velocity and projected radius vectors.
The circular orbit approximation allows us to compare 
the derived orbital parameters with observations in the inner
Galaxy, and to probe whether a circular orbit assumption is valid 
for the Arches. 

In a spherically symmetric potential, the gravitational
force only depends on the mass interior to the present 
galactocentric radius. While the potential in 
the inner galaxy is not spherically symmetric, this assumption is 
made in the concept of the enclosed mass vs.~radius relation providing 
the gravitational potential for previous simulations. As shown below,
this concept is too simplistic to explain the 3D space motion of the 
Arches cluster.

\begin{figure}[t]
\includegraphics[width=7.4cm]{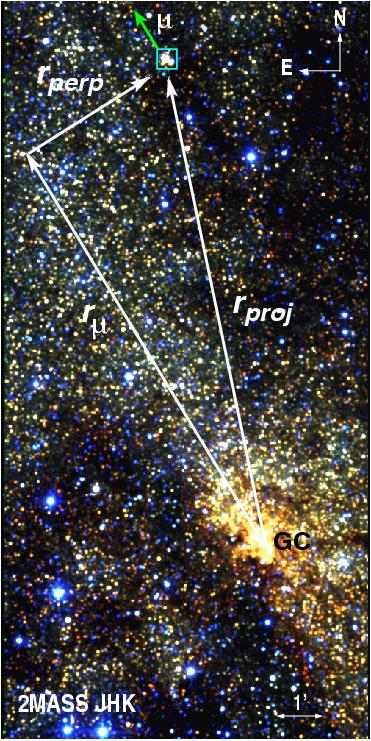}
\caption{\label{arch2mass}
The Arches cluster proper motion (green arrow) with respect to 
the GC is displayed on a 2MASS JHK color image. 
The bright band of stars extending from the GC into 
the direction of the radius vector \mbox{\boldmath$r_\mu$} indicates 
the Galactic plane, with the Arches at a projected height 
of about 10 pc above the disk. The cluster is moving almost
parallel to the plane, and away from the Sun.
The vectors illustrate the coordinate system defined 
to estimate the 3D distance of the cluster to the GC
under the assumption of a circular orbit.}
\end{figure}

\boldmath
For a circular, closed orbit, the absolute distance of the Arches
to the GC, $r_{GC}$, can be derived geometrically, 
using $v\cdot r$ = 0.
A convenient coordinate system can be defined from 
i) the direction along the proper motion vector $\mu$,
ii) the direction perpendicular to $\mu$ on the plane of the sky, and 
iii) the direction along the line of sight (see Fig.~\ref{arch2mass}).
As the 3D velocity of the cluster is defined by $\mu$ and the 
line-of-sight velocity, $v_{los}$, the velocity component perpendicular to $\mu$ 
is zero. The circular orbit equation thus becomes
\unboldmath

\begin{equation}
\mbox{\boldmath$v\cdot r$} = \mu r_{\mu} + v_{los} r_{los} = 0
\end{equation}

\noindent
where $r_{los}$ is the radius vector along the line of sight.
With $\mu = 212 \pm 29$ km/s, $r_{\mu} = r_{proj} cos\phi = 25.2 \pm 5.9$ pc,
where $\phi = {\rm arctan}(v_{los}/\mu) =18.4^o$ is the angle between 
the projected radius vector,
$r_{proj}$, and the proper motion vector on the plane of the sky,
and $v_{los} = +95 \pm 8$ km/s \citep{figer2002}, we obtain 
$r_{los} = - 56.4 \pm 25.5$ pc, pointing towards us.
The GC distance of the cluster in the circular orbit approximation 
would then be $r_{GC} = \sqrt{r_{proj}^2 + r_{los}^2} = 62 \pm 23$ pc.

As $r_{GC}$ was derived geometrically, no assumption about the 
enclosed mass was made. The enclosed mass required to obtain a 
circular, Keplerian orbit with an orbital velocity of $232 \pm 30$ km/s
at a radius of $62 \pm 23$ pc is $\sim (8\pm 3)\times 10^8\,M_\odot$.
However, from the modeled mass distribution in the inner Galaxy, the enclosed 
mass within $r_{GC} = 62$ pc is $2\times 10^8\,M_\odot$ 
\cite{laun2002}, where the mass contains both interstellar matter 
and stars as measured from near to far IR observations with IRAS and 
COBE/DIRBE.
This discrepancy in the inferred enclosed mass provides a first indication
that the Arches moves on a non-circular orbit.

\begin{figure}
\hspace*{-1.5cm}
\includegraphics[width=7cm,angle=90]{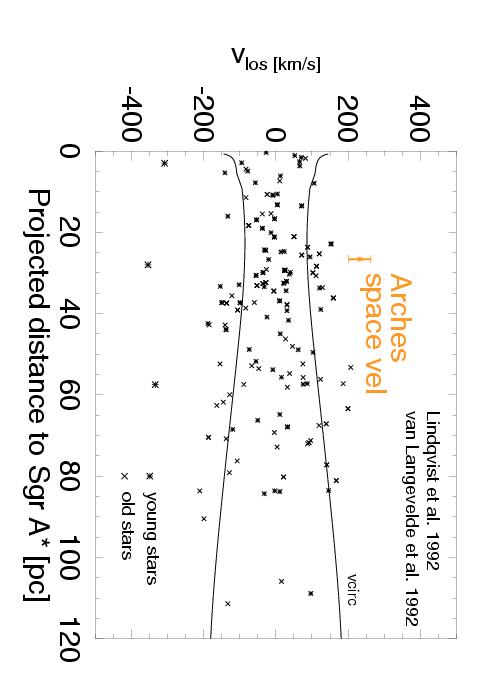}
\caption{\label{radvel}
Line-of-sight velocity of OH/IR stars observed by Lindqvist et al.~1992.
Enveloping lines are orbital velocities expected for circular,
Keplerian orbits derived from the enclosed mass (Launhardt et al.~2002).
The space motion of the Arches (red asterisk) is significantly larger 
than expected for a circular, Keplerian orbit at a projected radius
of 26 pc, implying that the cluster is either at a significantly larger 
absolute distance to the GC, or is not on a circular orbit, or both.}
\end{figure}

We can further probe the circular orbit approximation by 
comparing the measured 3D velocity of the cluster with 
observed and expected velocities in the inner Galaxy.  
In a sample of circularly orbiting objects along the line of sight, 
the orbital velocity can be measured as the maximum line-of-sight velocity 
observed at each projected radius.
Following van Langevelde et al.~(1992), we reproduce the line-of-sight velocities 
of a survey of OH/IR stars in the inner bulge conducted by  
Lindqvist et al.~(1992) in Fig.~\ref{radvel}, along with the 
expected maximum circular velocity at each projected radius 
derived from the enclosed mass determination by Launhardt et al.~(2002).
The circular velocities envelop most of the line-of-sight velocity measurements 
of bulge stars in Fig.~\ref{radvel}, suggesting that circular motion is 
a reasonable approximation for stellar orbits in the inner 120 pc.
Outliers with significantly larger velocities can be understood 
as stars on highly elongated orbits. 

The current orbital velocity of the Arches, $232 \pm 30$ km/s, lies
significantly above the orbital velocity of $v_{circ}= 86$ km/s
expected for a circular orbit at its projected distance of 
$r_{GC} = 26.6$ pc (cf.~Fig.~\ref{radvel}). 
The expected $v_{circ}$ increases to 120 km/s at $r_{GC} = 62$ pc, 
but remains almost a factor of two lower than the space velocity of 
the cluster. The large space velocity of the Arches places 
the cluster either at very large galactocentric radii,
where the Galactic rotation curve flattens ($r \ge 1$ kpc),
which is inconsistent with the cluster's suggested interaction 
with central molecular zone (CMZ) clouds (Lang et al.~2003, Wang et al.~2006),
or suggests a non-circular orbit. A non-circular orbit would also
explain the overestimated enclosed mass derived above.\footnote{%
Three OH/IR stars with similarly high
velocities of $\sim 200$ km/s are observed between projected radii
of 53 to 65 pc. These stars could have been ejected from 
close binary systems or dense clusters such as the Quintuplet or 
the Arches itself. While stars in the high source density of 
the inner Galaxy have short two-body relaxation timescales, 
a massive cluster is not easily deflected from a regular orbit. }

A circular cluster orbit in a spherically symmetric potential
is inconsistent with the measured enclosed mass
at the radius $r_{GC}$ and with observed circular velocities
in the distance range considered for the Arches.

\subsection{Relationship to gas cloud orbits: \newline 
            A scenario for the origin of the Arches}
\label{velsec}

\begin{figure}
\includegraphics[width=7.4cm]{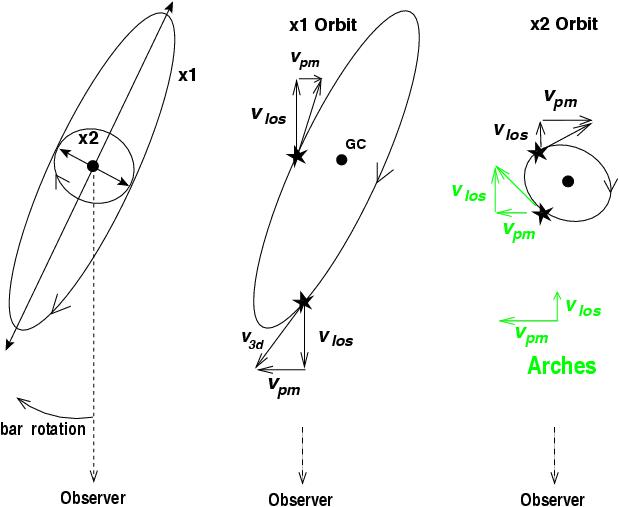}
\caption{\label{orbits}
Schematic representation of gas cloud orbits in the bar potential
in the inner Galaxy. The left panel displays the 
orientation of x1 and x2 orbits with respect to the Galactic bar
minor and major axes. The x1 orbital family reaches out to several 
kpc, while x2 orbits are confined to the inner 280 pc of 
the central bar. 
The middle panel displays proper motion and line-of-sight velocity 
vectors expected for objects on x1 orbits, which are
inconsistent with the measured Arches velocity vectors 
depicted at the lower right (green). 
The proper motion and line-of-sight velocity directions of objects on 
x2 orbits (right panel) are consistent with the observed Arches 
motion (green). However, as discussed in Sec.~\ref{orbsec},
the magnitude of the cluster velocity is too large for expected 
x2 orbital velocities.}
\end{figure}

In the central region of the Galaxy, the two-body relaxation time 
for a cluster with $M_{cl}=10^4\,M_\odot$ embedded in a medium of
stellar density
$\rho\sim\; 4\times 10^2\,M_\odot\,{\rm pc^{-3}}$ \citep{laun2002} 
is on the order of a few $10^9$ years (following Binney \& Tremaine 1987, 
eq. 8-71). 
With an age of only 2.5 Myr, the cluster velocity should therefore 
carry the imprint of the motion of its native cloud. A comparison 
of the cluster space motion with the characteristic motion 
of gas clouds in the inner Galaxy might shed light 
on its origin. 
In the central $l=\pm 2^{o}$ of the Galaxy, the outer envelop
of the CO emission in the longitude, line-of-sight velocity 
($l-v_{los}$) plane displays a ``parallelogram'' structure, 
which is interpreted as the reaction of the gas to a bar 
potential (Binney et al.~1991, Englmaier \& Gerhard 1999). 
Depending on their distance from the GC, long-lived gas clouds 
are suspected from these models to move on non-circular, closed orbits 
in the co-rotating reference frame of the bar, which belong to 
the x1 orbit family aligned with the major axis of the bar and the x2 
orbit family aligned with the bar's minor axis (see Fig.~\ref{orbits}). 
Cloud motion at large GC radii is dominated by x1 orbits with line-of-sight velocities 
up to 270 km/s outside $l > 2^{o}$ ($r_{GC} = 280$ pc), while inside the 
corresponding radius x2 orbits are expected to populate the inner spheroid. 

Dame et al.~(2001) observe line-of-sight velocities of CO gas up to 230 km/s 
close to $l=2^{o}$.
While the Arches velocity would be consistent
with the peak velocities observed outside $l=2^{o}$, for objects 
on x1 orbits at positive longitudes and with positive proper 
motion (towards increasing l), {\sl negative} line-of-sight velocities 
should be observed, which is not the case (Fig.~\ref{orbits}). 
The positive $v_{los}$ observed for the Arches agrees in sign with x2 orbit
simulations, but the high space velocity is inconsistent
with the maximum x2 orbital velocity of 120 km/s suggested 
by the Englmaier \& Gerhard bar model.
Thus, it appears that the cluster is not on one of these orbital 
families suggested for gas clouds in the inner Galaxy.

How can the cluster have gained such a high space velocity at such small 
GC distance?
Binney et al.~(1991) point out that gas clouds with substantial 
molecular components are only observed on the innermost x1 orbit
($r_{GC} \sim 280$ pc)
and the x2 orbits at even smaller radii, but not in the outer bar. 
They provide the intriguing interpretation that gas clouds on inner 
x1 orbits may collide with clouds on the outermost x2 orbits, where
the cloud density is higher. The subsequent fate of the clouds depends
on the impact angle. Close to the x2 apocenter, clouds might merge and
lose kinetic energy in the process. In a less direct collision, 
SPH simulations by Athanassoula (1988, reproduced in Binney et al.~1991, 
their Fig.~6) suggest that the x1 cloud will be 
dispersed into the x1-x2 transition zone, and collide with incoming
material on the opposite side. In both cases, cloud material will 
experience shock compression. Binney et al.~suggest that the collision
causes shock-creation of molecules, and the friction might allow the remnant
x1 cloud to shed angular momentum and spiral in toward a stable x2 orbit. 

If this scenario holds for sufficiently massive clouds, we can speculate
that the x1-x2 collision could trigger a 
starburst event, which would take place during the inspiral of the 
shocked cloud toward the region of stable x2 orbits. 
The emerging stellar cluster would naturally inherit the 
cloud velocity depending on the point along the x1-x2 orbit 
transition where the massive stars evaporate the native material.
Once the cluster emerged from its native cloud, it is unaffected 
by collisions with ambient clouds, and can lose energy solely by 
the slow process of two-body relaxation. At this point,
the cluster can sustain a significant velocity component from the 
higher x1 orbital velocity.
This scenario could explain both the apparent orientation of the 
Arches velocity vectors in agreement with x2 orbits, and the high 
velocity preserved from the x1 orbit transition.

Although the dynamical properties of the cluster are well explained,
this theory has a significant caveat. For the Arches to inherit the 
high x1 velocity, the x1 cloud would have to be more massive and dense
than the x2 cloud, which contradicts the distribution of dense clouds
observed in the CMZ today.
The present-day clouds on x1/x2 orbits representing the bar potential
are measured in CO, which traces gas at average densities 
$n_{H_2} \sim 10^3 {\rm cm^{-3}}$, suspected to be too low for stellar 
fragmentation. The denser components with $n_{H_2} \ge 10^4 - 10^5 {\rm cm^{-3}}$
capable of intense star formation are efficiently detected in the 
CS $J 1-0$ transition. While CS is 
correlated with many of the intense CO peaks \citep{bally1988}, 
CS clouds are more confined to the inner parts of the CMZ ($ r_{GC} < 200$ pc)
and display maximum {\sl line-of-sight} velocities $v_{los} < 120$ km/s. 
These velocities, if representative of the orbital velocities of dense clouds, 
are consistent with clouds on x2 orbits, but appear inconsistent with the 
large orbital velocity measured for the Arches. 
While the large present-day velocity of the Arches 
is suggestive of the collision scenario, we need to understand the initial 
conditions at the time of the starburst to constrain whether dense clouds 
could have been triggered to form the cluster, and in particular whether a large
orbital velocity at the present projected position can be acquired from these
initial conditions.
For this purpose, we perform 
a backwards integration of the cluster motion in the observed, non-spherically 
symmetric extended mass distribution in the inner Galaxy.

\section{The Arches motion in the GC potential}
\label{orbitsec}

The knowledge of the 3-dimensional space motion of the cluster
and the projected distance from the GC allow us to investigate possible orbits
of the cluster in the potential of the inner Galaxy depending on
the cluster's line-of-sight distance. 
The circular orbit considerations in Sec.~\ref{orbsec}
suggest that motion on a circular orbit in a spherically symmetric 
potential cannot explain the high 3D velocity of the cluster.
As the stellar density in the inner 
Galaxy changes dramatically with radius, we exploit the knowledge 
of the existing density of matter to approximate a more realistic 
gravitational potential. 
The cluster is evolved in this potential to yield
constraints on its present location with respect to the GC,
the possible position on its orbit, and its origin.\footnote{We use a 
leap-frog integrator originally developed by Joshua Barnes for a single 
logarithmic potential, which was extensively adapted to represent the 
gravitational components in the inner Galaxy. The basic version 
of the orbit integrator is publicly available from his lecture notes
(see http://www.ifa.hawaii.edu/~barnes/ for more information).}

\subsection{Modelling the potential in the inner Galaxy}
\label{modelsec}

Three components are suspected to contribute to the mass distribution
inside 200 pc, namely, the central black hole, the nuclear stellar cluster, 
and the thick nuclear disk comprised of both gas and stars. 
From their observed mid- to far-infrared luminosity distributions, 
Launhardt et al.~(2002) describe the density profiles of 
the nuclear stellar cluster as a spherically symmetric power-law decline,
and the nuclear disk as a combination of two  exponentials with additional
flattening in the latitude direction. 
Beyond 200 pc, the observed nuclear disk declines rapidly and the bar 
becomes dominant according to the observations. 
In the central few parsecs, the 
black hole is also an important contributor to the potential, and
is added as a point mass with $M=3.6\cdot 10^6\,M_\odot$.

As the only free variable of the cluster velocity and space coordinates is the distance 
along the line of sight, 
we define a sky coordinate system from the observer's perspective with $x$ the projected
distance from the GC along the Galactic plane, $y$ the line-of-sight distance, and $z$ the 
height above the Galactic plane. The line-of-sight distance
is defined such that zero corresponds to the distance of the GC from the sun. 
A positive line-of-sight distance implies the present cluster location is behind the GC.
Because the proper motion vector is oriented parallel to the Galactic plane, 
this coordinate system is similar
to the geometric system defined in Sec.~\ref{orbsec} to estimate the circular orbit,
where the $x$ axis is parallel to the proper motion vector, and the $y$
coordinate represents the unconstrained radius vector along the line of sight.

As the bar is inclined by $25^{\rm o}$  towards positive longitude
with respect to the line of sight \citep{rattenbury2007}\footnote{The bar 
rotation angle changes with the pattern speed of
$60 \pm 5$ srd/Gyr \citep{bissantz2003}, implying a rotation of $9^{\rm o}$ in 
2.5 Myr. On prograde orbits, the travel time of the cluster through the co-rotating 
bar potential becomes longer, while the travel time through the bar is shortened in 
retrograde orbits, causing orbits at large GC distances ($r_{GC} > 200$ pc)
to rotate in the same direction as the bar with respect to the line of sight. 
As the effect during the cluster lifetime is small and the most likely location 
of the cluster is inside 200 pc, we do not account 
for bar rotation in this model.}, while all other components are axisymmetric,
we convert the initial cluster position and velocity into the bar system, with $x_b$ along
the major axis and $y_b$ along the minor axis of the bar, while the height above the plane, 
$z$, is not modified. After evolving the cluster motion in the combined potential, the
output positions, velocities, and accelerations are converted back to the sky coordinate 
system.

\begin{figure}
\hspace*{-1.5cm}
\includegraphics[width=7cm, angle=90]{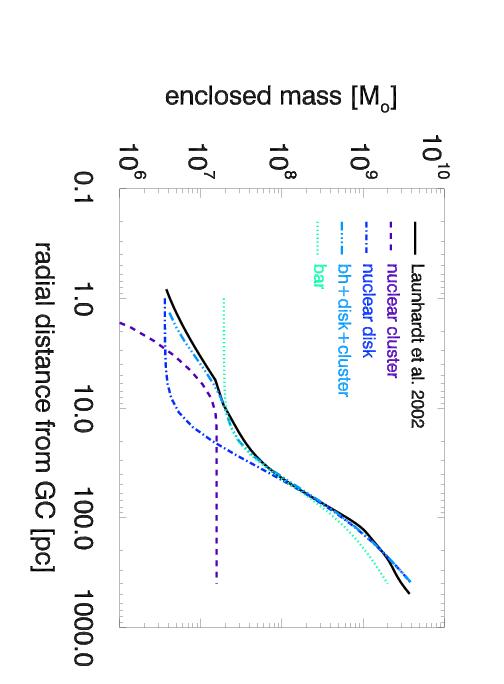}
\caption{\label{massfit} Fit to the observed enclosed mass (Launhardt et al.~2002, see 
also Lindqvist et al.~1992). In the central 200 pc, the gravitational potential
is represented by a spherical component, the nuclear stellar cluster, and an axisymmetric 
component flattened in galactic latitude, the nuclear stellar disk. The black hole is added 
as a point mass. The estimated enclosed mass from the combined potentials (bh+disk+cluster)
reproduces the enclosed mass very well. 
Beyond 200 pc, the potential is represented by an elongated bar.
The bar slightly underestimates the enclosed mass, because during the transition from the 
inner to the outer potential a continuous acceleration is enforced to avoid non-physical 
behaviour. }
\end{figure}

In order to calculate the combined gravitational potential at every position,
we use logarithmic potentials with different stretching parameters to represent
these components. The basic form of the adopted logarithmic potential is

\begin{equation}
\Phi = 0.5 v_0^2\, \log ( R_c^2 + \frac{x_b^2}{a^2} + \frac{y_b^2}{b^2} + \frac{z^2}{c^2} )
\end{equation}

\noindent
where $v_0^2$ represents the velocity on the flattened portion of the rotation
curve at very large radii, and $R_c$ the core radius. Both constants are free 
parameters in the fit to the observed mass distribution.
The stretching parameters $a$, $b$, and $c$ describe the flattening of the potential along 
the $x_b$, $y_b$ and $z$ axis, respectively, with respect to a spherically symmetric
potential where $a=b=c=1.0$. For the axisymmetric components representing the central 
cluster and thick disk this definition implies $a=b=1.0$.
In order to derive $v_0^2$ and $R_c$ for each component,
we spatially integrate the modeled density distribution corresponding 
to the logarithmic nuclear disk and cluster potentials to fit 
the observed enclosed mass in the central 200 pc 
(Launhardt et al.~2002, their Fig.~14). 
Additional constraints are given by the requirement 
that the acceleration, i.e., the first spatial derivatives of the potential, 
be continuous. The best mass fit under these 
conditions is shown in Fig.~\ref{massfit}, and the adopted
parameters for each component are summarized in Tab.~\ref{masstab}.

In our adopted model, the nuclear cluster is the dominant mass component 
between 2 and 7 pc, beyond which 
it is smoothly truncated to a constant mass, and the thick nuclear disk becomes 
the dominant component 
beyond 20 pc, while both contribute significantly to the mass at intermediate radii.
The sum of all three components, including the black hole, provides a very good 
approximation to the best available estimate of the enclosed 
mass in the inner 200 pc of the Galaxy. 

The transition between the inner components and the bar 
potential poses the largest problem for computational continuity. In order to ensure that
the acceleration vary smoothly and the energy be conserved along the orbit, the disk 
is transformed into an elongated bar potential in the transition zone between 150 and 250 pc.
Technically, a smooth variation of the potential is achieved by multiplying a sigmoid function
of the form $1/(1+e^{-r/f})$ to the stretching parameters.
Here, $r = r_{GC} - r_{trans}$ is the difference between the radial distance from the GC 
and the midpoint of the transition region, $r_{trans} =200$ pc, and 
$f$ is a scaling factor that defines the steepness of the transition. The sigmoid function 
is 0 for radii significantly below the transition radius $r_{trans}$, and 1 for radii 
significantly above this radius. The scaling factor is chosen to be $f=15$, which ensures
a gentle transition over the full range of radii, $150 < r_{trans} < 250$ pc.
For a smooth transition between the nuclear disk and the bar potential, 
we therefore modify the x-stretching and z-stretching parameters from the axisymmetric disk with
$a_{disk}=1.0$ and $c_{disk} = 0.71$ to become the elongation of the bar, $a_{bar}=c_{bar}=0.75$ 
(Binney et al.~1991)\footnote{%
Binney et al.~1991 derived an elongation of 0.75 by modeling line-of-sight velocities of 
gas clouds in the inner Galaxy. More recent determinations of the bar potential 
from red clump stars suggest a significantly stronger bar elongation with a 
stretching factor of $\sim 0.3$ (Bissantz \& Gerhard 2002, Rattenbury et al.~2007). 
The minimum stretching parameter allowed in the analytic density solution 
of a logarithmic potential
is 0.71 to ensure a positive density at all positions. A more realistic, elongated bar 
potential requires the full numerical integration of the density distribution to obtain 
the cluster orbit, which is beyond the scope of this paper. 
We therefore adopt the $q$ parameter derived from cloud velocities. This choice only affects
cluster orbits beyond $r_{GC} = \pm 200$ pc.}, as follows \\
\\
$a(r) = a_{disk} - (a_{disk}-a_{bar})/(1.+{\rm exp}(-r/15.)) $ \\
$c(r) = c_{disk} + (c_{bar}-c_{disk})/(1.+{\rm exp}(-r/15.)) $ \\
\\
while b remains 1.0 in the bar reference frame. At any given position, the local 
conversion factor between the disk and the bar is taken into account in the calculation
of the potential and the acceleration, and the subsequently derived velocity and 
position vector. Beyond ensuring mathematical continuity, the smooth transformation of 
the thick disk into the bar can be interpreted as one continuous
potential in the inner Galaxy, which is known to be disk-like in the central region and 
bar-like at larger radii, while the exact shape of the potential combining the two 
components is unknown.

\subsection{Constraints on the cluster orbit}

The family of orbits of the Arches cluster was calculated over 
a range of line-of-sight distances to the GC from -500 to 500 pc.
The absolute distance is not well constrained, even if formation 
scenarios for GC clusters suggest the cluster to be within the 
inner 200 pc. The most stringent evidence for a location close to the GC
of the Arches are i) the mean foreground extinction of $A_V = 27$ mag, 
which is consistent with the line-of-sight extinction towards the GC 
(Nagata et al.~1995, Cotera et al.~1996), 
and ii) the suggested interaction of the cluster with clouds in the central 
molecular zone in the inner $\sim 200$ pc of the Galaxy (Wang et al.~2006,
Lang et al.~2003). We note that i) provides only a weak 
constraint on the 
Arches distance due to the strong spatial variability of the extinction 
observed in the GC environment.
For example, in the central parsec of the nuclear stellar cluster, extinction 
values between $24 < A_V < 34$ mag are measured (Schoedel et al.~2007).
We thus decided to cover the largest plausible
distance range taking into account these constraints.

\begin{figure*}
\includegraphics[width=7cm, angle=90]{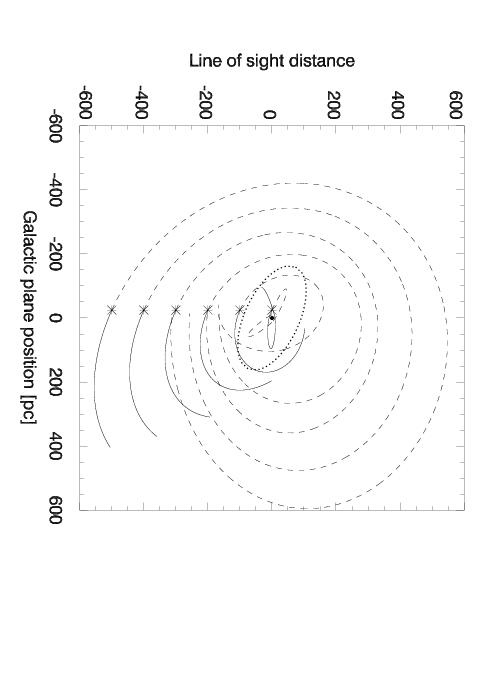}
\includegraphics[width=7cm, angle=90]{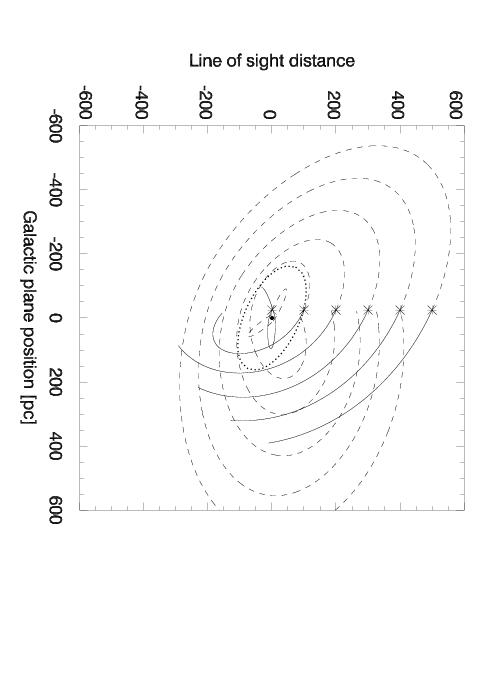} \\
\includegraphics[width=7cm, angle=90]{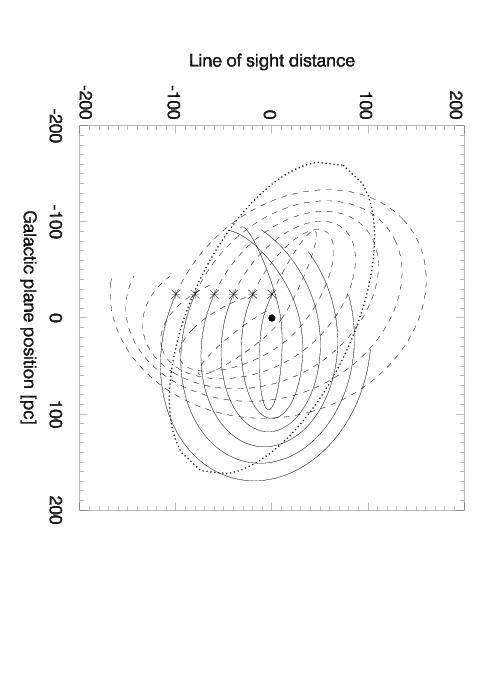}
\includegraphics[width=7cm, angle=90]{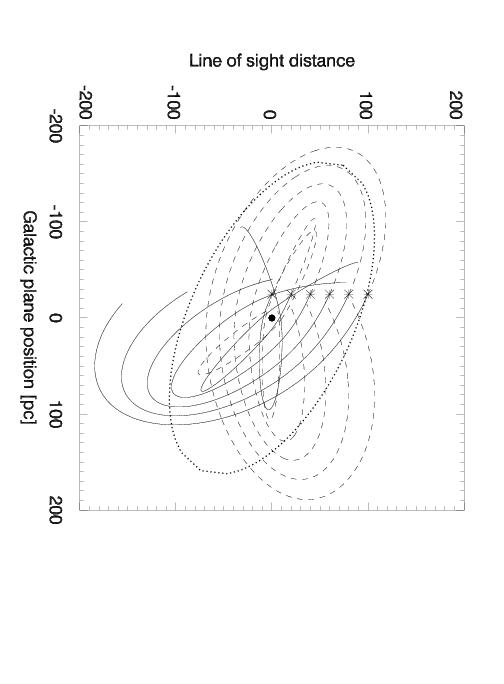}
\caption{\label{selorbits} Selected orbits in the combined central potential.
The orbits are viewed from above the Galactic plane, with the x-axis displaying 
the projected distance from the Galactic center in the direction of the Galactic
plane, and the y-axis displaying the line-of-sight distance to the observer. 
The asterisks mark the present location of the cluster for various line of 
sight distances.
The GC is located at the origin.  The solid lines represent the backwards integration 
of the Arches velocity to the possible cluster origin 2.5 Myr ago. The dashed lines 
show the extrapolated future orbit to one full $360^o$ revolution around the GC.
The nuclear stellar cluster and disk, including the black hole, are relevant 
out to 200 pc, beyond which the bar potential dominates. The location of the outermost
x2 orbit for a bar inclination of $25^o$ is indicated by the dotted line 
(model from Bissantz et al.~2003).}
\end{figure*}

\begin{figure}
\includegraphics[width=7.4cm]{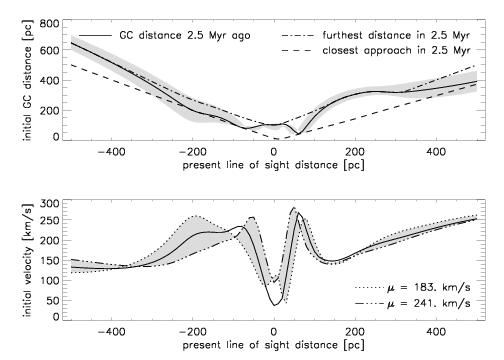}
\caption{\label{init} Initial orbital parameters of the Arches cluster at the suspected time of 
birth $\sim2.5$ Myr ago. The top panel displays the initial GC distance, which 
would have been the distance of the cluster-forming cloud at the time when the starburst
occured, while the bottom panel shows the initial cluster velocity. 
Both the furthest and closest approach of the cluster to the GC depend almost linearly 
on its present-day GC distance. The asymmetries are due to the inclined bar potential.
The uncertainty of $\pm 29$ km/s in the proper motion as propagating into the orbital parameters
are indicated by the underlying grey area, which was derived from limiting models 
with $\mu_{min} = 183$ km/s and $\mu_{max} = 241$ km/s.}
\end{figure}

\begin{figure}
\includegraphics[width=7.4cm]{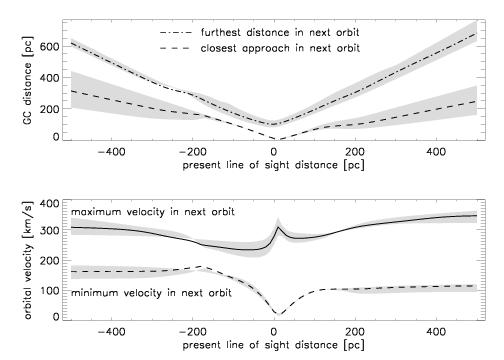}
\caption{\label{orbfig} Extrapolated next orbit of the Arches cluster around the GC.
The top panel shows the closest and furthest distance from the GC during the next 
orbit. The bottom panel displays the minimum and maximum orbital velocities the 
cluster can attain on its orbital path. The grey area indicates uncertainties
in the orbital simulations from the uncertainty in the proper motion of $\pm 29$ km/s.}
\end{figure}

\begin{figure}
\includegraphics[width=7.4cm]{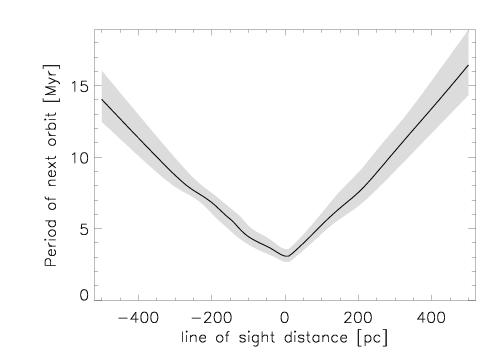}
\caption{\label{period} Period vs. line-of-sight distance.
The extrapolated period of the next full orbit of the Arches cluster
in the asymmetric GC potential. As in Figs.~\ref{init} and \ref{orbfig}, 
the grey area indicates the uncertainty in the orbital parameters from 
the one sigma uncertainty in the cluster velocity.}
\end{figure}

Orbits were calculated in steps of 10 pc along the line of sight
both forwards and backwards in time. Backwards orbits are considered
over the 2.5 Myr since the formation of the cluster to constrain
possible initial conditions.
As most of the allowed cluster orbits in the asymmetric potential
are not closed, we define one orbit as a change in position angle of $360^{\rm o}$.
In the following discussion, we will refer to parameters originating 
2.5 Myr ago as {\sl initial conditions} assumed to be close to the 
time of cluster formation, and to parameters projected from the cluster's 
present position along the first full orbit into the future as {\sl the next orbit}.
The results are presented in Figs.~\ref{selorbits}-\ref{period} and 
Tabs.~\ref{orbtabback}, \ref{orbtabfor}.

The uncertainty in the orbital parameters was estimated with two modified orbital
simulations, in addition to the standard model represented in Tab.~\ref{masstab}.
In order to probe the effects of the shape of the potential, one set of 
orbits was simulated in a more spherical potential, with $a=c=0.8$ for the bar
and $c=0.8$ for the disk. Note that our standard model $c=0.71$ is the flattest
logarithmic potential available without obtaining negative densities in the 
central region. In the modified potential, the apocenter and pericenter distances,
minimum and maximum velocities along the initial and next orbit, as well as 
the period, were only marginally different from the standard model. We thus 
do not show the results here, as they are indistinguishable from
the parameters shown in Figs.~\ref{selorbits}-\ref{period}. In a second test, 
the cluster proper motion, the largest velocity component contributing most 
of the uncertainty in the cluster motion, was modified to 
$\mu \pm \sigma_\mu = 212 \pm 29$ km/s, from 183 km/s to 241 km/s. The largest 
deviation in both apocenter/pericenter distances and orbital velocities occurs 
if the cluster's present line-of-sight distance is larger than 200 pc.
In general, the orbital characteristics are very similar when a larger or
smaller present-day proper motion is assumed. These test cases confirm that
the derived orbits are not very sensitive to details in the logarithmic 
potential or the uncertainties in the cluster velocity.

Sample orbits of the cluster in the Galactic plane are shown in Fig.~\ref{selorbits}. 
Only the motion in the plane is shown, which is superposed upon a small oscillation of 
less than 50 pc above and below the plane.
For clarity, orbits originating from a cluster position in front of the GC (left panels) 
are shown separately from present-day positions behind the GC
(right panels). It is noteworthy that, due to the positive velocity along the 
line of sight, all orbits where the cluster is presently behind the GC are retrograde
to the general motion of stars and gas clouds in the bar potential, 
and all orbits where the cluster
is presently in front of the GC are prograde. If the cluster formed from a cloud on a 
prograde orbit, it has to be located in front of the GC today. 
For all present-day locations inside $r_{GC} < 100$ pc, the high velocity
suggests that the cluster recently passed the pericenter of its orbit and now 
moves towards apocenter.

The approximate shape and location of the outermost x2 orbit as 
suggested by the recent bar model of Bissantz et al.~(2003) is
shown as a dotted ellipse in Fig.~\ref{orbits} (top panels).
For a cluster location in front of the GC today,
a line-of-sight distance of 100-200 pc is most consistent 
with a formation locus near the outermost x2 orbit as suggested 
by x1-x2 cloud collision scenarios. Such a formation scenario is less 
likely for orbits where the cluster is found behind the GC at present.
At line-of-sight distances larger than 200 pc, the cluster origin seems 
unrelated to the x1-x2 transition zone.
Triggered cluster formation from dispersed x1 clouds
entering the x2 orbital zone is also feasible if the cluster is 
inside the inner 100 pc of the Galaxy, both for line-of-sight distances 
in front of and behind the GC. 
Although the direct comparison of the cluster formation locus and the 
outermost x2 orbit is complicated by the uncertainty of $\pm 0.5$ Myr
in the cluster age and thus its exact origin, as well as by the extent of the 
outermost x2 orbit populated by dense clouds and the uncertainty in 
the bar orientation angle, the suggested scenarios support the origin 
of the Arches inside the central 200 pc. 

The initial conditions of the molecular cloud at the time when the starburst
occured $\sim2.5$ Myr ago are depicted in Fig.~\ref{init}. 
The general trend suggests that the distance of closest and furthest approach to 
the GC in the past 2.5 Myr both scale linearly with the present-day distance 
of the cluster from the GC.
The asymmetries originate from the passage through the inclined bar potential. 
When the cluster is presently located in front of the GC, it approaches the bar
and its path was not strongly influenced by the bar 
potential in the past 2.5 Myr. If, on the other hand, the cluster is 
located behind the GC, it had to pass through the bar recently
and experienced the increased gravitational forces from the central bar potential.
In the latter case, the Arches is now moving away from the 
major axis of the bar and will not encounter the region of highest density 
in the bar in the near future.
The absolute minimum of the distance to the GC occurs when the Arches is on a 
radial orbit. Given the orientation of the measured velocity vector, the Arches 
would have to be at a line-of-sight distance of about 20 pc behind the GC today to be 
on a radial orbit. On nearly radial orbits, the Arches would have come 
closer than 10 pc to the GC in the past, thereby likely suffering intense 
tidal stripping. Such tidal stripping should be detectable in a strong tail 
of tidal debris distributed along the cluster orbit. Once stars are not bound 
to the cluster potential anymore, the relevant two-body relaxation time shrinks 
to the stellar two-body relaxation time of $\sim 10^7$ yr in the nuclear cluster. 
One might therefore expect that tidally stripped stars suffer dynamical 
interactions with nuclear cluster sources, which may lead to a rapid dispersal
of the tidal tail during future passages through the nuclear cluster.
Kim \& Morris (2003) simulate the dispersal of a cluster with an initial 
core density of $10^5-10^6\,M_\odot/{\rm pc^{-3}}$ starting at 30 pc
on a highly elongated orbit. Such a cluster can deposit at most 5-10\% of 
its stellar population at radii $r < 5$ pc inside the nuclear cluster, 
and practically no stars in the central parsec. Thus, even if the Arches 
is on a radial orbit, we do not expect to find Arches stars near the supermassive
black hole (SMBH).

Of all orbits, the only ones which can bring the cluster into the inner 10 pc 
of the GC in the past 2.5 Myr, require a present-day
GC distance of $r_{GC} < 33$ pc, very close to its observed projected distance.
Any larger distance suggests that the estimated total stellar mass of $\sim 10^4\,M_\odot$ 
\citep{stolte2002} is very close to the initial cluster mass. 
As inspiraling cluster models require initial masses of $M_{cl} \ge 10^6\,M_\odot$ 
from a distance of 30 pc (Kim \& Morris 2003), the orbital analysis
observationally confirms the numerical prediction (Kim et al.~2000)
that the initial mass of the Arches was never substantial
enough for inspiral into the central few parsecs.

Characteristics of the projected future orbit of the Arches are shown in Fig.~\ref{orbfig}.
The top panel shows the distance of closest and furthest approach, suggesting that,
regardless of its origin, the cluster can approach the GC to within 200 pc during 
the next few Myr.
The approximately linear dependence of closest approach upon present line-of-sight distance 
implies that without energy loss, the cluster cannot get much closer to the black hole 
during the next orbit than at its present position. The only exception is the 
innermost orbit with $r_{los} = 0$, where the cluster has passed beyond pericenter 
and will reach a closer pericenter position only a few pc from the SMBH 
after half a revolution.
The bottom panel  in Fig.~\ref{orbfig} displays the minimum and maximum velocity 
during the next orbit.
Both the minimum GC distance and the maximum velocity show pronounced extrema
at $r_{los} = 20$ pc, where the cluster follows an almost radial trajectory
towards and away from the GC.
The spike in the maximum velocity curve corresponds to the point of closest
approach, where the cluster passes the black hole within a few parsecs.

The projected orbit into the immediate future of the cluster provides additional
clues on the likelihood of cluster survival. Fig.~\ref{period} shows the period
of the next orbit. If the cluster presently resides in the inner 100 pc, 
it will have completed about one orbit in its lifetime (see also Fig.~\ref{selorbits}, 
lower panels). 
At larger GC distances, the section of the orbit covered becomes accordingly smaller
(see also Fig.~\ref{selorbits}, upper panels).
At line-of-sight distances $r_{los} > 100$ pc, the cluster remains at distances $r_{GC} >$ 100 pc
during the next orbit
if it is presently located in front of the GC, and at $r_{GC} >$ 65 pc if it is 
located behind the GC, suggesting 
that tidal forces will only intensely influence the cluster if it is found to be 
inside the inner 100 pc. 
This ambiguity shows the need for a precise distance measurement to
the Arches cluster.

At a present-day line-of-sight distance of -10 pc, the {\sl mean distance} 
to the GC on the next orbit is about 70 pc, only marginally 
higher than the 62 pc estimated from the circular orbit approximation. Nevertheless,
the orbital period of 3.2 Myr is twice as long as the 1.6 Myr estimated for a 
circular orbit
in a spherically symmetric potential from the parameters derived in Sec.~\ref{orbsec}, 
showing the need to account for a realistic potential in the inner Galaxy.
At line-of-sight distances of 300 pc and more, the cluster needs about 15 Myr to 
complete one orbit. 

In summary, the Arches cluster has completed at most one orbit during its lifetime,
and can pass through the inner 10 pc only if the cluster's galactocentric radius 
is very close to its projected distance of 26 pc.
These constraints imply that the Arches cluster cannot deposit young stars into the
GC, supporting the prediction from N-body simulations by Kim \& Morris (2003) 
that the dense cores of clusters at large projected radii ($r_{gc} > 10$ pc)
are unlikely to reach the central parsec even on highly elongated orbits.
The orbital analysis incorporating the high orbital motion of the cluster
thus provides the first observational evidence 
that Arches-like clusters are neither progenitors of the nuclear cluster, 
nor contribute significantly to the young population in the central parsec.

\subsection{Comparison with dense clouds in the CMZ}

\begin{figure*}
\includegraphics[width=12cm,angle=90]{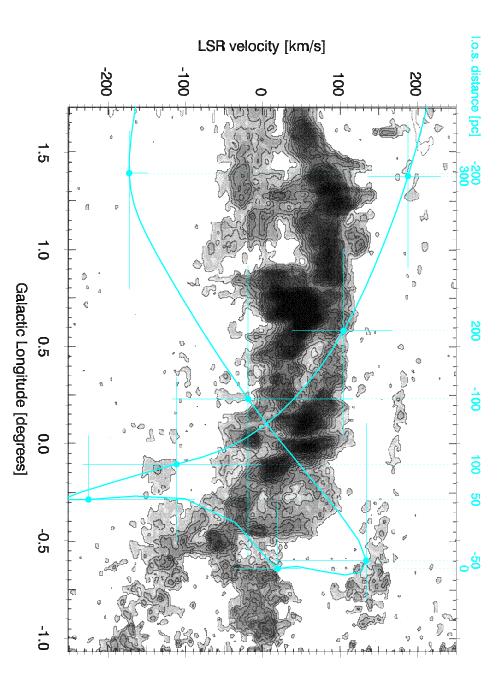}
\caption{\label{csmap} CS $J=1-0$ ($l, v_{los}$) map of dense molecular clouds 
in the CMZ as observed by Tsuboi et al.~(1999). 
The position of the Arches integrated backwards along its 
orbit for 2.5 Myr, are shown as blue dots, and the solid lines are error bars
allowing for an age uncertainty of $\pm 0.5$ Myr. 
The present-day line-of-sight distance of the cluster is labeled on the top axis, and
the dotted lines connect each data point with the corresponding present-day
line-of-sight distance. Initial conditions for line-of-sight distances $> 200$ pc 
are indicative due to the neglected rotation of the bar.}
\end{figure*}

As pointed out at the end of Sec.~\ref{velsec}, the dense, massive, 
molecular clouds in the CMZ are found inside 200 pc with moderate 
velocities of up to 120 km/s. With knowledge of the cluster's initial
conditions, the ($l$, $v_{los}$) location of the native clouds of the 
Arches at the time of its birth can be entered into the CS map of 
dense molecular gas observed in the GC today (Fig.~\ref{csmap}, see 
Fig.~2 in Tsuboi et al.~1999). Error bars allow for an age uncertainty
of $2.5 \pm 0.5$ Myr for the Arches \citep{najarro2004}.

Assuming that the spatial distribution of clouds 2-3 Myr ago
was similar to the CMZ observed today,\footnote{The maximum x2
line-of-sight velocity does not change strongly with the rotation 
of the bar. See Binney et al.~1991 for effects of changes in the inclination
angle on the ($l$, $v_{los}$) diagram.} the extrapolated initial conditions 
for the Arches cluster coincide with the location of dense, massive clouds 
($M \sim 10^6\,M_\odot$, Miyazaki \& Tsuboi 2000) for most of the line-of-sight 
distances. Large line-of-sight distances above 200 pc in front of or behind 
the GC appear less likely because of the expected high initial velocities, 
$v_{los} > 170$ km/s, significantly larger than the observed cloud velocities, 
$v_{los} \le 120$ km/s. At present-day line-of-sight distances larger 
than 300 pc from the GC, the orbital analysis suggests that initial velocities 
would have to be $v_{los} > 300$ km/s, which is unusually large for a star-forming 
cloud in the CMZ.
At small radii, the initial velocities depend strongly on the present age.
In the inner 100 pc, essentially all initial velocities are possible, indicating 
that the initial velocity is not well constrained at small GC distances.
In summary, the comparison with the space-velocity distribution of dense, massive
clouds in the CMZ indicates a preferred formation locus inside the inner 200 pc 
for the Arches. The possible orbital paths of the cluster suggest that the Arches
is in this case also within $\pm 200$ pc of the GC today (see Fig.~\ref{init}).

\section{Summary and conclusions}
\label{sumsec}

We measure the proper motion of the Arches cluster 
with respect to the surrounding field to be $212 \pm 29$ km/s.
Combination with the line-of-sight velocity yields the 3D space motion
of the cluster relative to bulge stars, $v_{3D}=232 \pm 30$ km/s. 
The high orbital velocity is inconsistent with a circular orbit 
in a spherically symmetric GC potential. 

In order to constrain the orbit, we evolve the cluster in a more 
realistic potential following the observed mass distribution in the 
inner Galaxy. We obtain the cluster's initial position and velocity 
as a function of its present-day line-of-sight distance. Comparison 
with CS observations suggests that the initial position and line-of-sight velocity 
of the Arches coincide with x2 orbits of dense and massive 
($\sim 10^5 - 10^6\,M_\odot$, Miyazaki \& Tsuboi 2000)
molecular clouds if the cluster formed inside $r_{GC} < 200$ pc,
in which case it should still reside within 200 pc today. 
The high orbital velocity contradicts the recently suggested 
formation of the Arches as a rejuvenated globular cluster (Lin \& Murray 2007),
as this scenario requires a small relative motion of the cluster with 
respect to GC clouds for an extended period of time.

In the asymmetric potential of the inner Galaxy, the Arches completed at most 
one orbit during its lifetime,
and came into the central 10 pc only if it is presently located closer than 
33 pc from the GC, close to its projected distance. These constraints 
restrict the amount of material possibly tidally stripped from the cluster, 
and render it highly unlikely that star clusters with Arches properties suffer 
sufficient energy loss to migrate inwards during the lifetime of the massive 
stars and contribute to the young population in the central parsec of the GC.
This implies that with the Arches and Quintuplet clusters we are not observing
progenitors to the GC comoving groups, providing observational confirmation 
of inspiral simulations (Gerhard 2001, Kim \& Morris 2003).
This supports the origin of the nuclear cluster from a substantially more massive 
cluster at small radii a few million years ago, or via in-situ star formation in 
the immediate GC environment.

Given the young age of the cluster and its long orbital relaxation 
timescale in the inner Galaxy, the large space velocity
should be inherited from the natal cloud.
When comparing the cluster orbital motion with line-of-sight velocities
of gas clouds in the inner Galaxy using models of orbits in the 
dominating bar potential, we find that the 3D cluster motion is 
inconsistent with the closed x1 and x2 orbits suggested for gas clouds.

Although ample molecular material exists in the inner 200 pc which could 
harbour the formation of clusters like the Arches, the dense, massive clouds 
have maximum line-of-sight velocities
below 120 km/s, consistent with x2 orbital velocities, but significantly 
lower than the present-day orbital velocity of the cluster. 
In the standard cloud fragmentation scenario, a dense, massive cloud 
must have existed on a high-velocity orbit 2.5 Myr ago
to generate the cluster. 
In addition, if a significant fraction of the dense clouds on x2 orbits would
form starbursts, more clusters like the Arches or Quintuplet should be 
observed. The low number of clusters
in comparison to clouds on x2 orbits and the high proper motion of the Arches
indicate that the starburst event that could give rise to a massive compact
cluster might have been triggered by a cloud-cloud collision.

Intense cloud-cluster interaction is observed from the X-ray
bow shock of the Arches with its ambient medium by Wang et al.~(2006),
who suggest that a cloud-cloud collision could 
also have triggered the formation of the cluster.
Following Binney et al.~(1991), the starburst could have been 
triggered while the native cloud of the Arches 
collided on the boundary between x1 and x2 orbits 
in the inner bar, or from dispersed cloud material crossing the
x1-x2 transition region. The backwards integration of the cluster
orbit is consistent with this formation locus if the cluster resides
in the inner 200 pc.
However,  for the Arches to inherit the large orbital velocity
from an x1 cloud, this model requires a massive, high-density cloud orbiting 
on an x1 orbit 2.5 Myr ago, while clouds sufficiently dense to be 
triggered into an intense starburst event are only observed in the x2 orbital 
zone today. Alternatively, a collision with an infalling high-velocity 
cloud (e.g., Crawford et al.~2002) could also trigger star 
formation and propel the cluster on a highly eccentric orbit. 
The fact that the Arches moves predominantly 
parallel to the Galactic plane and has a vertical velocity of 
only $10$ km/s indicates that the cluster did not originate
from a collision with a vertically infalling cloud.
If both the Arches (2.5 Myr ago) 
and Quintuplet (4 Myr ago), as well as Sgr B2 were triggered 
by cloud collisions, this would indicate a substantial 
cloud collision in the inner $\sim 200$ pc approximately 
every 2 Myr. 

A cloud collision was suggested as the trigger for 
massive star formation in Sgr B2 by Hasegawa et al.~(1994),
who interpret the velocity structure observed in $^{13}$CO
as a dense clump colliding with a massive molecular cloud 
of lower density. These authors investigate two scenarios of 
cloud collisions near the Galactic center, i) a collision 
between two CMZ clouds, one of which moves on an elongated 
orbit in the bar, and ii) an infalling cloud originating 
above the Galactic plane. The first scenario can originate
from a cloud on an x1 orbit moving into higher density clouds
on x2 orbits, as suggested above. However, Hasegawa et al.~(1994) 
crudely estimate the rate for such collisions to be as low as 
one collision per 200 Myr, which could not explain the occurence
of the Arches, Quintuplet, and Sgr B2 at the same time.
These authors point out that the rate of infalling clouds
is not well constrained, and they do not provide an estimate for 
the second scenario. 
Bally et al.~(1988) identify a low-density molecular structure 
that is inclined by up to $40^o$ out of the Galactic plane, 
and point out that material on such inclined orbits has to 
cross the CMZ clouds in the plane twice along one orbital path,
leaving ample opportunity for cloud-cloud interactions.
Alternatively, the Arches could have an exceptionally high 
velocity, and might have formed in a different mode than 
its neighbours. Currently, neither the collision rate nor
the spontaneous starburst rate in the GC are well constrained. 
The observations of 3D space motions of the neighboring clusters 
will shed additional light on their origin.

Our measurement of the cluster motion will 
constrain dynamical evolution models in a realistic
bar and nuclear cusp potential including energy loss 
along the orbit, which is the required 
next step to quantify the tidal mass loss of the cluster.
Finally, the proper motion selection of cluster members 
provides us with a powerful tool to remove the field
contamination, and derive the tidal structure and unbiased 
stellar mass function of the Arches cluster, which will be 
discussed in a forthcoming paper.

\acknowledgements
The authors wish to thank the anonymous referee for thoughtful 
comments on the manuscript and in particular the cluster formation 
scenarios. We thank Brad Hansen, Mike Rich and Hong Sheng Zhao for 
insights on the dynamical properties in the inner Galaxy and orbital
considerations, and Ralf Launhardt for kindly providing his enclosed 
mass measurements and density distributions, and acknowledge fruitful 
discussions with Mike Muno and Claire Max.
WB acknowledges support by a Julian Schwinger Fellowship at UCLA.
This work would not have been possible without the intense effort 
and dedication of the Keck LGS-AO staff. We are deeply greatful
for their support enabling these observations. The W. M. Keck Observatory
is operated as a scientific partnership among the California Institute
of Technology, the University of California, and the National Aeronautics
and Space Administraction. The Observatory was made possible by the generous
financial support of the W. M. Keck Foundation. The authors wish to recognize
and acknowledge the very significant cultural role and reverence that the 
summit of Mauna Kea has always had within the indigenous Hawaiian community.
We are most fortunate to have the opportunity to conduct observations from 
this mountain.
This work was supported by NSF grant AST 04-06816 and by the
Science and Technology Center for Adaptive Optics,
managed by the University of California at Santa Cruz under
cooperative agreement AST 98-76783.

\clearpage


\begin{table}
\caption{\label{masstab} Parameters for nuclear cluster, disk and bar in the logarithmic potential
approximation}
\begin{tabular}{lccccc}
potential & a & b & c & $R_c$ & $v_0$ \\
 &  & & & pc & km/s \\
\hline
nuclear stellar cluster & 1.0 & 1.0 & 1.0 & 2.0 & 98.6 \\
nuclear stellar disk    & 1.0 & 1.0 & 0.71 & 90. & 190. \\
bar                     & 0.75 & 1.0 & 0.75 & 90. & 190. \\
\end{tabular}
\end{table}


\begin{table}
\caption{\label{orbtabback} Orbital parameters in the past 2.5 Myr vs. 
present-day line-of-sight distance}
\begin{tabular}{ccccc}
 $r_{los}$ & $r_{GC}$ init & $v$ init & $r_{GC}$ min & $r_{GC}$ max \\
pc & pc & km/s & pc & pc \\
\hline
-500. & 647 & 132 & 501 & 647 \\
-400. & 516 & 128 & 401 & 518 \\
-300. & 367 & 144 & 301 & 389 \\
-200. & 197 & 211 & 197 & 267 \\
-100. & 107 & 229 & 103 & 172 \\
-50. & 86 & 186 & 56 & 127 \\
0. & 100 & 36 & 11 & 101 \\
50. & 60 & 227 & 31 & 141 \\
100. & 156 & 171 & 69 & 194 \\
200. & 305 & 167 & 142 & 305 \\
300. & 317 & 203 & 221 & 317 \\
400. & 345 & 231 & 296 & 401 \\
500. & 391 & 253 & 372 & 501 \\
\end{tabular}
\end{table}

\begin{table*}
\caption{\label{orbtabfor} Orbital paramters of the next full orbit vs. present
line-of-sight distance and absolute GC distance, $r_{GC}$, for each orbit.}
\begin{tabular}{ccccccc}
$r_{los}$ & $r_{GC}$ & $r_{GC}$ min & $r_{GC}$ max & vmin & vmax & Period \\
  pc  &  pc  &  pc  &  pc  &  km/s  &  km/s  &  Myr  \\
\hline
-500. & 501 & 313 & 620 & 161 & 310 & 14.0 \\
-400. & 401 & 257 & 493 & 162 & 304 & 11.4 \\
-300. & 301 & 203 & 369 & 163 & 289 & 8.7 \\
-200. & 202 & 166 & 287 & 174 & 262 & 6.8 \\
-100. & 103 & 100 & 172 & 139 & 237 & 4.4 \\
-50. & 57 & 56 & 127 & 104 & 234 & 3.8 \\
0. & 27 & 10.0 & 101 & 28 & 278 & 3.1 \\
50. &  57 & 31 & 141 & 60 & 272 & 4.0 \\
100. & 103 & 68 & 194 & 96 & 276 & 5.3 \\
200. & 202 & 99 & 304 & 103 & 309 & 7.5 \\
300. & 301 & 145 & 432 & 109 & 327 & 10.4 \\ 
400. & 401 & 197 & 556 & 112 & 341 & 13.4 \\
500. & 501 & 247 & 683 & 114 & 347 & 16.4 \\
\end{tabular}
\end{table*}

\clearpage



\clearpage


\clearpage


\clearpage


\clearpage


\end{document}